\documentclass{PoS}
\usepackage{amsmath, amsthm, amssymb}
\usepackage{stmaryrd}
\usepackage[utf8]{inputenc}

\title{Dual gravity with $R$ flux from graded Poisson algebra}

\ShortTitle{Dual gravity, $R$ flux and graded P.B.}

\author{\speaker{Eugenia Boffo}\\ 
        Jacobs University Bremen\\
        E-mail: \email{e.boffo@jacobs-university.de}}

\author{Peter Schupp\\
       Jacobs University Bremen\\
               E-mail: \email{p.schupp@jacobs-university.de}}

\abstract{ We suggest a new action for a ``dual'' gravity in a stringy $R$, $Q$ flux background. The construction is based on degree-$2$ graded symplectic geometry with a homological vector field. The structure we consider is non-canonical and features a curvature-free connection. It is known that the data of Poisson structures of degree $2$ with a Hamiltonian correspond to a Courant algebroid on $TM \oplus T^{*}M$, the bundle of generalized geometry. With the bracket for the Courant algebroid and a further bracket which resembles the Lie bracket of vector fields, we get a connection with non-zero curvature for the bundle of generalized geometry. The action is the (almost) Hilbert-Einstein action for that connection.}

\FullConference{Corfu Summer Institute 2019 "School and Workshops on Elementary Particle Physics and Gravity" (CORFU2019)\\
		31 August - 25 September 2019\\
		Corfù, Greece}

\begin{document}

\section{Introduction}

In the standard approach to gauge theories, the interactions are due to the minimal coupling of the conserved current with the matter field. An alternative viewpoint is offered by Poisson geometry and a Hamiltonian function. Deforming the Poisson brackets is another equivalent method for introducing interactions with the gauge fields. To realize a gravitational theory for the metric tensor, the Poisson structure should be capable of accommodating  the symmetric metric tensor. This is indeed possible if graded variables with odd parity are brought into play: then the exterior algebra of functions of these variables is actually symmetric. Graded Poisson brackets with a Hamiltonian are rich in structure: via derived brackets, they have been shown to correspond to (Courant or Lie) algebroids on vector bundles \cite{Roy02} \cite{YKS03}. The gauge fields will hence be incorporated into the higher algebra. Switching to the algebroid perspective shifts the focus away from the matter fields, but manages to put in a dominant position the free theory for the metric and the relevant gauge fields, which can be readily written down, for example as the Hilbert-Einstein action of a connection. 

In this article the aforementioned $2$-graded Poisson algebra approach is used with a metric $G^{-1}$ for cotangent space rather than tangent space, together with a bivector $\Pi$. These $2$-tensors are motivated from string theory. $G^{-1}$ can be interpreted as the effective metric perceived by the open strings, while $\Pi$ can be intended as a noncommutativity parameter for the string endpoints on a brane \cite{SW99}. We aim at building a ``dual'' gravity action, for $G^{-1}$, in the background of the field strength of $\Pi$. Connotating $G^{-1}$ as ``dual" metric should not be confused with the dual graviton of higher spins theory and M-theory. What is meant here with ``dual" is motivated by the open-closed strings relation, as used in \cite{And12}, \cite{And13} and \cite{Blu13}. The action and its construction should shed new light on the stringy $R$ and $Q$ fluxes, which are discussed in the context of non-geometry in string compactification and non-associativity of D-branes.

\section{Deformed graded Poisson brackets}
The basic idea is the following: We would like to enhance the canonical graded symplectic form of $T^{*}[2]T[1]M =: \mathcal{M}$,
\[
\omega_{0} = dx^{i} \wedge dp_{i} + \dfrac{1}{2} \eta^{i}{}_{j} d\chi_{i} \wedge d \theta^{j},
\]
to a deformed symplectic form, allowing a more general metric than the standard pairing $\eta $. This can possibly lead to a connection acting on the space of the linear functions of degree $1$, $\mathcal{O}_{1}(\mathcal{M})$, whose curvature is zero. By doing so, we want to employ the closed-open string relation 
\[
\left(g \pm B\right)^{-1} = G^{-1} \pm \Pi.
\]
Let us begin with discussing a local open set of $\mathcal{M}$ and the degree (or grading) we assigned to its coordinates. The open set is $4d$-dimensional, and a chart is made up with coordinates $\{x^{i}\}^{d}_{i=1}$ on $M$, of degree $0$ (these are standard coordinate for a manifold), $\{x^{i}, \theta^{i}\}$ for $T[1]M$, where the degree of $\theta$ is $1$ and hence $\theta^{i}\theta^{j} = -\theta^{j} \theta^{i}$. The remaining coordinates are the canonical conjugate pair of the former coordinates: $\{\chi_{i}, p_{i}\} $ of degree $1$ and $2$ respectively, and since the $\chi$'s have odd degree their parity is odd too. Sometimes we will label the collection $(\chi_{i}, \theta^{j})$ with $\xi_{\alpha}$, the Greek indices running from $1$ to $2d$.
 
The deformed symplectic form $\omega$ is given by
\begin{equation}
\omega = dx^{i} \wedge dp_{i} + \frac{1}{2} d \chi_{j} \wedge d\theta^{j} + G^{ij} d \chi_{i} \wedge d \chi_{j} + dx^{k} \, \partial_{k} \left(G^{-1}-\Pi\right)^{ij} \chi_{i} \, \wedge d\chi_{j}\,,
\end{equation}
where $G^{-1}$ and $\Pi$ are in general $x$-dependent.
Its total degree is $2$. It is easy to be convinced that 
\[
\omega = \omega_{0} + d A, \quad A = \left(G^{-1}-\Pi\right)^{ij} \chi_{i} d \chi_{j},
\]
i.e.\ $\omega$ and $\omega_{0}$ belong to the same cohomology class. The transformation from $\omega_{0}$ to $\omega$  is generated by the Moser vector field \cite{moser}
\[
X = \left(G^{-1} - \Pi\right)^{ij}\chi_{i} \dfrac{\partial}{\partial \theta^{j}}.
\]
The central object of our considerations will be the Poisson brackets corresponding to $\omega$. Let us explicitly write them down, using $f \in C^{\infty}(M) \cong \mathcal{O}_{0}(\mathcal{M})$, $U = X + \sigma$ and $V = Y + \kappa$ in $\mathcal{O}_{1}(\mathcal{M})$, and $\upsilon(x):=\upsilon^{i}(x) p_{i}$:
\begin{align}
\{\upsilon(x), f(x)\} &= \upsilon(x) . f(x), \notag  \\
\{U(x), V(x)\} &= \begin{pmatrix} 0 & \, & \iota_{X} \kappa \\ \iota_{Y} \sigma & \, & 2G^{-1}(\sigma, \kappa) \end{pmatrix}, \label{pbd} \\ 
\{\upsilon(x), V(x)\} &= \bar{\nabla}_{\upsilon} V(x) = \upsilon(x) . V(x) + \upsilon^{k}(x) \kappa_{i} \partial_{k} \left(G^{-1}+\Pi\right)^{ij} \chi_{j}. \notag
\end{align}
All omitted brackets are zero, and the ``dot'' denotes the action of the preceding element on the $C^{\infty}(M)$ function that follows. Notice that when a non-constant metric is introduced, internal consistency (such as validity of the Jacobi identity) brings up the metric connection $\bar{\nabla}$.

There is a differentiable structure attached to this symplectic structure. It is encoded in the graded vector field
\begin{align}
\text{Q} = & \, p_{i} \frac{\partial}{\partial \chi_{i}} + \left[ \left(G^{-1} + \Pi\right)^{ij} p_{j} -\left(\theta^{j} - \chi_{k} \left(G^{-1}-\Pi\right)^{kj} \right) \partial_{j} \left(G^{-1} + \Pi\right)^{im} \chi_{m} \right] \frac{\partial}{\partial \theta^{i}}  \notag \\ \, & \, + \left(\theta^{i} - \chi_{j} \left(G^{-1} - \Pi\right)^{ji} \right) \frac{\partial}{\partial x^{i}} . \label{qvf}
\end{align}
$\text{Q}$ preserves the symplectic structure, i.e.\;$\mathcal{L}_{\text{Q}} \omega=0$. In graded symplectic geometry (under the mild condition that the sum of the degrees of $\text{Q}$ and $\omega$ is non-zero) this means that it is also Hamiltonian~\cite{Roy02}. As such, it is homological, i.e.\;$\text{Q}^{2} =0$. With the Hamiltonian function $\Theta$ that is defined by
\[
\text{Q} = \{ \cdot, \Theta\},
\]
the condition $\text{Q}^{2} = 0$ is equivalent to the \emph{classical master equation}:
\begin{equation}
\{\Theta, \Theta\} =0.
\end{equation}
The Hamiltonian function for \eqref{qvf} is 
\begin{equation}
\Theta = \left(\theta^{i} - \chi_{j} \left(G^{-1}-\Pi\right)^{ji} \right) p_{i}. \label{HAM}
\end{equation}
When both a graded symplectic structure and a differentiable structure are present the situation becomes even richer: one can get new brackets defining a higher algebraic structure for some vector bundles, via the derived brackets of the Poisson algebra with $\text{Q}$.

The infinitesimal gauge symmetries for \eqref{pbd} which are also degree preserving can be shown to integrate to diffeomorphisms and $O(d,d)$ transformations. How about the higher infinitesimal gauge symmetries of these symmetries? We will be interested in the invariances for the degree $2$ generators which are quadratic in the degree $1$ coordinates, such as $B \in \Lambda^{2}T^{*}[1]M$ or $\beta \in \Lambda^{2} T[1]M$. We know these should be given by $\text{Q}$-exact degree $2$ functions. In the actual case under examination, they originate from
\begin{equation}
\varrho = \varrho_{i}(x) \left( \theta^{i} - \left(G^{-1}+\Pi\right)^{ij} \chi_{j}\right).
\label{varrho}
\end{equation}
Then $\text{Q} \varrho$ corresponds to
\begin{equation}
\text{Q} \varrho = \left(\theta^{l} - \chi_{m} \left(G^{-1}-\Pi\right)^{ml}\right) \left(\partial_{l} \varrho_{i}(x)\right) \left( \theta^{i} - \left(G^{-1}+\Pi\right)^{ij} \chi_{j}\right) . \label{Qvarr}
\end{equation}

In the rest of the article it would be relevant to attach a ``relaxed'' differentiable structure to the canonical Poisson structure $\omega_{0}^{-1}$:
\begin{equation}
\{p_{i}, x^{j}\} = \delta_{i}^{j}, \quad \{\chi_{i}, \theta^{j}\} = \delta_{i}^{j}.
\label{pbc}
\end{equation}
The graded vector field we would like to consider for this matter is
\begin{equation}
\text{\v{Q}} = p_{i} \dfrac{\partial}{\partial \chi_{i}} + \left(G^{-1}+\Pi\right)^{ij} p_{j} \dfrac{\partial}{\partial \theta^{i}}. \label{vQ}
\end{equation}
$\text{\v{Q}}$ trivially squares to zero, but fails to preserve the symplectic form $\omega_{0}$. Hence it is not Hamiltonian for the symplectic form with Poisson brackets \eqref{pbc}. It might nevertheless be useful to understand in which way the Hamiltonian function $\Theta$ of total degree $3$ in \eqref{HAM} fails to solve the classical master equation, when the Poisson structure is canonical:
\[
\{ \Theta, \Theta \} = -2\left(G^{-1}\right)^{jk} p_{j} p_{k}, \quad \text{with the Poisson brackets \eqref{pbc}}.
\]
The deviation from the master equation is indeed particularly nice and deserves some further attention: the theory ceases to be topological and the dynamics of a test particle is given by the geodesic equation.
\begin{align*}
\dot{x}^{i} =&  \dfrac{1}{4} \{ x^{i}, \{\Theta, \Theta\}\} = \left(G^{-1}\right)^{ij} p_{j}, \\
\ddot{x}^{i} =& \dfrac{1}{4} \{ \dot{x}^{i}, \{\Theta, \Theta \}\} = \Gamma^{kl}{}_{j} G^{ji} \, \dot{x}_{k} \dot{x}_{l} ,
\end{align*}
where $\Gamma$ is the Christoffel symbol for $G$: $\Gamma^{kl}{}_{j} := \dfrac{1}{2} \left(-\partial_{j} G^{kl}+ 2G_{ji} G^{(k \vert m} \partial_{m} G^{\vert l) j}\right)$. However the relativistic dynamics shall not be the main focus here: we will rather use $\{,\}$ \eqref{pbc} and $\text{\v{Q}}$ \eqref{vQ} for other scopes in the next section.

\section{Higher algebra structure}

To the graded phase space, extended with degree $1$ velocities and its Hamiltonian vector field, one can associate a Courant algebroid on the bundle $TM \oplus T^{*}M \equiv E$ \cite{Roy02}. A Courant algebroid is the collection of four objects: a vector or principal bundle over $M$, a map from the total space to~$TM$, brackets for the sections and a fiber-wise symmetric bilinear form. They mutually have to respect some axioms \cite{Gua03}. The vector bundle inherits a fiber-wise metric $\mathcal{G}$ from the graded Poisson structure \eqref{pbd}:
\begin{equation}
\mathcal{G}( U(x), V(x)) = \begin{pmatrix} 0 & \iota_{X} \kappa \\ \iota_{Y} \sigma & 2G^{-1}(\sigma, \kappa) \end{pmatrix}. \label{mathG}
\end{equation}
The isometry group for the metric $\mathcal{G}$ is $O(d) \times O(d) \rtimes \exp(\beta)$, where $\beta \in \Lambda^{2} T^{*}[1]M \subset \mathfrak{o}(d,d)$ is exponentiated to be in the component of $O(d,d)$ connected to the identity. We will reserve the pointed brackets $\langle , \rangle \equiv \eta$ for the standard pairing between forms and vector fields. The anchor map $\rho : \Gamma(E) \mapsto \Gamma(TM)$ descends from a derived bracket with $f \in \mathcal{O}_{0}(\mathcal{M})$ and $V \in \mathcal{O}_{1}(\mathcal{M})$:
\begin{equation}
\{\text{Q} f, V\} = -\{ \text{Q}V, f\} = \rho(V)f = \left( Y^{i} + \left(G^{-1}+\Pi\right)(\kappa)^{i} \right) \partial_{i} f(x).
\label{anchor}
\end{equation}
Clearly the map is not just a projector onto $TM$.
The (non-antisymmetric) algebroid bracket $[,]$ is a derived bracket in which the arguments are two functions in $\mathcal{O}_{1}(\mathcal{M})$:
\begin{equation}
[U(x),V(x)] = \{\text{Q} U(x), V(x)\}. \label{derdorf}
\end{equation}
Therefore, due to the peculiar Poisson structure with a connection \eqref{pbd}, in the algebroid bracket the connection $\bar{\nabla}$ appears:
\begin{align}
[U(x), V(x)] = & \, \bar{\nabla}_{\rho(U)} V(x) - \bar{\nabla}_{\rho(V)} U(x) + \mathcal{G}( \bar{\nabla}_{\tilde{\rho}(\cdot)} U(x), V(x)) \notag \\
\, =& \, \left( X^{i} + \left(G^{-1} + \Pi\right)(\sigma)^{i} \right) \bigg(\partial_{i} V(x) + \kappa_{j} \partial_{i} \left(G^{-1}+\Pi\right)^{jk} \chi_{k} \bigg) \notag \\
\, & - \left(Y^{i} + \left(G^{-1} +\Pi\right)(\kappa)\right)^{i} \bigg(\partial_{i} U(x) + \sigma_{j} \partial_{i} \left(G^{-1}+\Pi\right)^{jk} \chi_{k} \bigg) \notag \\
\, & + \mathcal{G}\bigg(\left(\theta^{i} - \chi_{j} \left(G^{-1} - \Pi\right)^{ji} \right) \left(\partial_{i} U(x) + \sigma_{j} \partial_{i} \left(G^{-1}+\Pi\right)^{jk} \chi_{k} \right), V(x)\bigg). \label{der-D}
\end{align}
In the first row, $\tilde{\rho}: \Gamma(E^{*}) \mapsto \Gamma(TM)$. The quadruple $(TM \oplus T^{*}M, [,] \,\eqref{der-D}, \rho \,\eqref{anchor}, \mathcal{G} \,\eqref{mathG})$ can be shown to respect the defining axioms of a Courant algebroid. In particular, the axioms hold as long as $\Theta$ solves the classical master equation.
 
Using the isomorphism $(\chi_{i}, \theta^{i}) \cong (\partial_{i}, dx^{i})$, expression \eqref{der-D} is seen to partly contain the Dorfman bracket $[,]_{\text{D}}$, the standard example of bracket for a Courant algebroid,
\[
[U,V]_{\text{D}} = [X,Y]^{k} \, \partial_{k}  + \left( X^{i} \partial_{i} \kappa_{k}  - Y^{i} \partial_{i} \sigma_{k} + \langle V, \partial_{k} U\rangle \right) \, dx^{k}. 
\]
We should expect this since the metric $\mathcal{G}$ deviates from the $O(d,d)$-invariant pairing $\langle , \rangle$ only on the $1$-forms, and when the anchor is applied to an $E$-section which consists purely of a vector field it simply projects it.

A new connection $\nabla$, metric for $\mathcal{G}$ and with totally antisymmetric torsion, can be brought into play if an antisymmetric and $\mathbb{R}$-linear bracket of $E$-sections $\llbracket, \rrbracket$ is subtracted from the algebroid bracket, as proposition $4.4$ in \cite{Boffo19} ensures. We argue in our companion paper \cite{Boffo20} that such a bracket, that we called generalized Lie bracket, is a derived bracket too, namely of the Poisson structure \eqref{pbc} with the derivation $\text{\v{Q}}$ \eqref{vQ}, which is not a homological vector field. The antisymmetric bracket corresponds to
\begin{align}
\llbracket U, V\rrbracket &=: \{\text{\v{Q}} U,V \} - \{\text{\v{Q}} V, U\} \label{gen-exlb} \\
\, & = \bigg(\left(X^{i} + \left(G^{-1}+\Pi\right)(\sigma)^{i}\right) \partial_{i} V^{\alpha}(x) - \left(Y^{i} + \left(G^{-1}+\Pi\right)(\kappa)^{i}\right) \partial_{i} U^{\alpha}(x) \bigg) \xi_{\alpha}.
\end{align}
The generalized Lie bracket \eqref{gen-exlb} can be shown to respect the Jacobi identity as a consequence of the properties of the graded Poisson brackets and of the $\text{\v{Q}}$ derivation. It reduces to ask that 
\[
\{\{\text{\v{Q}} U, \text{\v{Q}} V\}, W\} - \{\text{\v{Q}}\{\text{\v{Q}}U,V\}, W\} \vert_{\text{antisymmetrized in} \; U,V} = 0.
\]
This is true if the map $\rho$ \eqref{anchor} is a homomorphism of $\llbracket, \rrbracket$ with the Lie bracket of vector fields,
\[
\rho(\llbracket U, V\rrbracket ) = [\rho(U), \rho(V)]_{\text{Lie}}.
\]

Having \eqref{der-D} and \eqref{gen-exlb} at our disposal we can then write down the closed expression for the connection $\nabla$, 
\begin{equation}
\mathcal{G}\left( \nabla_{\tilde{\xi}^{\alpha}} U, V\right) \xi_{\alpha} = [U,V]^{\alpha} \xi_{\alpha} - \llbracket U,V\rrbracket^{\alpha} \, \xi_{\alpha}, \label{propo}
\end{equation}
where $\tilde{\xi}^{\alpha}$ is the dual coordinate. The computation gives:
\begin{align*}
\mathcal{G}\left( \nabla_{\tilde{\xi}^{\alpha}} U, V\right) \xi_{\alpha} =& \, \left(X^{i} + \left(G^{-1}+\Pi\right)(\sigma)^{i}\right) \kappa_{j} \partial_{i} \left(G^{-1}+\Pi\right)^{jk} \partial_{k} \\
\, & - \left(Y^{i} + \left(G^{-1}+\Pi\right)(\kappa)\right)\sigma_{j} \partial_{i} \left(G^{-1}+\Pi\right)^{jk} \partial_{k} \\
\, & + \mathcal{G} \left( \theta^{i} - \chi_{j}\left(G^{-1}-\Pi\right)^{ji}\right) \left( \partial_{i}U(x) + \sigma_{j} \partial_{i} \left(G^{-1}+\Pi\right)^{jk}\chi_{k}, V \right).
\end{align*}
A connection on cotangent space $\widetilde{\nabla}: \Gamma(T^{*}M) \mapsto \Gamma(TM) \otimes \Gamma(T^{*}M)$, metric w.r.t.\;$G^{-1} \in \vee^{2} TM$, can also be defined. The existence of this rather unusual notion of differentiation is ensured by proposition $4.4$ of \cite{Boffo19} and by the non-degeneracy of $\mathcal{G}$ \eqref{mathG} restricted to $T^{*}M$. Consider plugging just $1$-forms in the brackets on the LHS of \eqref{propo} -- it will yield the following\footnote{Hatted elements are supposed to not be subjected to derivation.}
\begin{align}
\widetilde{\nabla}_{\zeta} \sigma =& \, \dfrac{1}{2} G_{jm} \bigg( \left(G^{-1}+\Pi\right)(\sigma)^{i} \partial_{i}\left(G^{-1}-\Pi\right)(\hat{\zeta})^{m} + \left(G^{-1}+\Pi\right)(\zeta)^{i} \partial_{i}\left(G^{-1}+\Pi\right)(\hat{\sigma})^{m}\notag \\
\, & \, + \left(G^{-1}+\Pi\right)^{mi} \partial_{i} \left(G^{-1}+\Pi\right)(\hat{\sigma}, \hat{\zeta}) + \left(G^{-1}+\Pi\right)(\zeta)^{i} \left(\partial_{i} \sigma_{j}\right) \, \bigg) \, dx^{j} . \label{consame}
\end{align}
The connection is interesting on its own right; it encompasses the standard definition of an affine connection since the direction along which the derivation is taken stems from a $1$-form, subsequently anchored to $TM$. The connection coefficients are reproduced here:
\begin{align*}
\tilde{\Gamma}^{ki}{}_{j} dx^{j} = \dfrac{1}{2}G_{jm} \bigg(&\, \left(G^{-1}+\Pi\right)^{kl} \partial_{l} \left(G^{-1}+\Pi\right)^{im} +\left(G^{-1}+\Pi\right)^{il} \partial_{l} \left(G^{-1}+\Pi\right)^{mk} \\ \, & -\left(G^{-1}+\Pi\right)^{ml} \partial_{l} \left(G^{-1}+\Pi\right)^{ik} \bigg) \, dx^{j}.
\end{align*}
This shows that $\widetilde{\nabla}$ naturally encodes a dual graviton (in the sense of dual gravity as explained in the introduction) and the derivatives on the bivector $\Pi$:
\begin{equation}
Q_{i}{}^{jk}:= \partial_{i} \Pi^{jk} , \quad R^{ijk} :=3! \, \Pi^{[i \vert l} \partial_{l} \Pi^{\vert jk]}
\label{QR}
\end{equation}
In string theory, these are the local expressions, for zero $H \in H^{3}(M, \mathbb{Z})$, of the fluxes that show up after T-dualizing a background with isometries and with a non-zero unit of $H$ flux. They are globally non-trivial if $\Pi$ patches as $\Pi_{(a)} =  \Pi_{(b)} + \delta \Pi_{(ab)}$,
\begin{equation}
\delta \Pi^{ij}_{(ab)} =  \left(G^{-1}+\Pi\right)^{il} \partial_{l} \alpha_{k} \left(G^{-1} -\Pi \right)^{kj},
\quad \text{see} \; \eqref{Qvarr}
\label{gauge-Pi}
\end{equation}
In the current setting $Q$ and $R$ are not strictly motivated by T-duality. They are certainly complementing a new gravitational theory, where they source the torsion of a metric connection for an essentially inverse metric $G^{-1}$; similarly to \cite{And12}, we recover that they transform as connection coefficients. We would hence like to get the curvature scalar of the connection for dual gravity with a bivector, and the corresponding invariant action. On this matter, we will report on our companion paper \cite{Boffo20} where, despite having a different graded symplectic form and Hamiltonian, the connection there coincides with \eqref{consame}.

\section{Dual gravity action}

The commutator of the covariant derivative minus the covariant derivative of the generalized Lie bracket is a curvature tensor, $\text{Riem}: \bigotimes^{3} \Gamma(T^{*}M) \mapsto \Gamma(T^{*}M)$:
\begin{equation}
\text{Riem}(\zeta, \varepsilon, \sigma) := \left[ \widetilde{\nabla}_{\zeta} , \widetilde{\nabla}_{\varepsilon}\right] \sigma - \widetilde{\nabla}_{\llbracket \zeta, \varepsilon \rrbracket} \sigma.
\end{equation}
The generalized Lie bracket has the correct properties, antisymmetry and the Leibniz rule, to match with the commutator of the covariant derivatives. Taking the partial trace of the first slot of $\text{Riem}$ and the open slot, leads to the analogue of the Ricci tensor of standard differential geometry, $\text{Ric} \in \bigotimes^{2} \Gamma(TM)$. As our basic tensor that gives rise to the deformation of the graded Poisson algebra is the combination $\left(G^{-1}+\Pi\right)$, the natural contraction of the Ricci tensor to a curvature scalar must involve the same non-symmetric combination, with indices lowered by $G$. As such, also the antisymmetric terms of $\text{Ric}$ are retained in the calculations. Thus,
\begin{equation}
\text{Riem}_{i}{}^{jik} G_{jl} \left(G^{-1}-\Pi\right)^{lm} G_{mk} =: \text{R},
\end{equation}
is the formula for the curvature scalar that we are going to employ. Noting that the connection coefficients which are due to the anchor applied to $G^{-1}$ build up the torsion-free part of the connection, i.e.\;the Levi-Civita connection $\nabla^{L.C.}_{G}$ of this setup, they will give rise to a Ricci scalar for $G^{-1}$. The rest is arranged into 
\begin{align}
\sqrt{\det{G^{-1}}} \, \text{R} =& \bigg(\text{R}_{G} - \dfrac{1}{12} R^{2} + \dfrac{3}{4} \left( Q_{m}{}^{jn} \left(G^{pm} Q_{p}{}^{ml}\right)\right) G_{lj} G_{nk} \notag \\
\, & - \left( \dfrac{1}{6} Q_{s}{}^{jn} + \dfrac{2}{3} G_{rs} G^{jp} Q_{p}{}^{nr} \right) R^{ksl} G_{jl} G_{nk} \bigg) \sqrt{\det{G^{-1}}}, \label{act}
\end{align}
where we tacitly integrated by parts, using the relation between the partial derivative of a half density accompanied with a vector and the divergence of the vector itself:
\[
\rho^{ij}\partial_{j} \left(\sqrt{\det{G^{-1}}}  w_{i} \right) = \sqrt{\det{G^{-1}}} \nabla^{L.C. \; i}_{G} w_{i} \equiv \sqrt{\det{G^{-1}}} \widetilde{\nabla}^{i} w_{i},
\]
as the antisymmetric connection coefficients do not contribute to the divergence.

Equation \eqref{act}, integrated on the manifold $M$, constitutes already the action functional that we were looking for. It incorporates ``dual'' graviton degrees of freedom, due to the metric $G^{-1}$, the field strength of the bivector $\Pi$ as well as $Q$ and $R$ interactions. By construction it is diffeomorphism invariant. However that is not all; it enjoys also a more hidden invariance. Gauge transformations of the bivector \eqref{gauge-Pi} are a genuine symmetry of \eqref{act}. To check this we can rely on the derived formalism and the degree-$2$ Poisson structure(s). Since the derived Dorfman bracket \eqref{derdorf} is invariant by construction ($\delta \Pi$ is a $\text{Q}$-exact $2$-vector), it remains to see whether $\llbracket ,\rrbracket$ is left unchanged by the transformation or not. But this is easy: with $\varrho$ in \eqref{varrho} and $\text{\v{Q}}$ in \eqref{vQ} one can immediately conclude:
\[
\text{\v{Q}}\varrho = 0.
\]

\section{Discussion and conclusions}

In this short article we presented an action for ``dual'' gravity with $Q$ and $R$ fluxes \eqref{act}. The starting point was a deformed graded Poisson structure together with a simple Hamiltonian function \eqref{HAM}. The ansatz for the deformation that accomplishes this particular task was simply a redefinition of the degree $1$ coordinates for $T[1]M$,
\[
\theta^{i} \mapsto \theta^{i} + \left(G^{-1}+\Pi\right)^{ij} \chi_{j}.
\]
Generally speaking, we hence believe that dg-manifolds are pretty suitable, via derived brackets construction and via the proposition of \cite{Boffo19}, to obtain connections for the bundle of generalized geometry. Furthermore the gauge freedoms of these connections are automatically fixed. The Poisson brackets we began with offer the foundations for deformation quantization. 

The action establishes a new viewpoint on a ``dual'' graviton, and it includes also an antisymmetric tensor $\Pi \in \mathfrak{X}^{2}(M)$. The quest for a geometric action with the stringy fluxes is not completely new, however. A quite similar action was suggested in \cite{And12} upon development of the differential geometry of cotangent space (for a manifold where standard coordinates are doubled with the winding modes). The action for $10$-dimensional supergravity with fluxes of \cite{And13} is very close to ours, the connection coefficients being anyway slightly different. The gauge symmetries of the bivector that the authors suggested in (4.33) are exactly ours \eqref{Qvarr}. We thus provided a neat derivation of the symmetry from arguments due to dg-symplectic manifolds. Other actions of this type were the outcome of the research of \cite{Blu13} and \cite{Wat15}. In the former the differential geometry of $T^{*}M$, with the Lie algebroid there, were deployed to provide a Levi-Civita connection and an invariant $3$-tensor, for the inverse of the $B$ field or for the more complicated $\Pi$ we also used. In the latter the authors could get gravity with $R$-flux from a Poisson manifold via derived brackets: but these, in the degree $1$ case, are Lie algebroid brackets. Here we have studied the degree $2$ case instead. 

From the mathematical viewpoint, we see some important achievements here. First of all the differential geometry of $TM \oplus T^{*}M$ is developed and new formulas for the torsion and the curvature tensor are suggested. The deformed graded Poisson algebra \eqref{pbd} is rather general and a further study is still at hand. It is anyway an applied example, and thus it can be generalized even more. Let us point out that the Hamiltonian function we selected is quite specialized. The overall Hamiltonian function is also available:
\[ \Theta = \xi_{\alpha} \tilde{\rho}^{\alpha i} p_{i} + \dfrac{1}{3!} C^{\alpha \beta \gamma} \xi_{\alpha} \xi_{\beta} \xi_{\gamma}. \]
The first term on which we focused exclusively, is analogous to a Dirac operator. The second twists the derived brackets, which are brackets for the Courant algebroid. In the current article we replaced the twisting of the bracket with a non-holonomic basis, and in this way were able to obtain local expressions of $Q$ and $R$ \eqref{QR} as connection coefficients. Lastly, we found quite intriguing that the Hamilton equations, due to the canonical Poisson structure with the Hamiltonian for the deformed case, manages to reproduce the correct dynamics of a particle probe moving in a curved space. Such dynamics is the geodesic equation, where parallel transport is performed via the Levi-Civita connection for $G^{-1}$. 

\section{Acknowledgments}

We are grateful to the RTG 1620 ``Models of Gravity'' for financial support. We want to warmly thank the organizers of the conference on ``Recent developments in Strings and Gravity'' for the wonderful program of the conference.

\end{document}